\documentclass[useAMS,usenatbib]{mn2e}

\usepackage{graphicx}
\usepackage{amssymb}
\usepackage{amsmath}
\pdfoutput=1

\def\beqa{\begin{eqnarray}}
\def\eeqa{\end{eqnarray}}

\def\beq{\begin{equation}}
\def\eeq{\end{equation}}

\def\d{{\bf d}}
\def\x{{\bf x}}
\def\hx{{\bf{\hat x}}}
\def\n{{\bf n}}
\def\A{{\bf{A}}}
\def\N{{\bf{N}}}
\def\M{{\bf{M}}}

\def\B{{\bf{B}}}
\def\W{{\bf{W}}}
\def\m{{\bf{m}}}
\def\u{{\bf{u}}}

\def\v{{\bf{v}}}
\def\w{{\bf{w}}}
\def\I{{I}}
\def\cI{{I}}
\def\Sv{{\bf{S}}}
\def\T{{\bf{T}}}

\def\O{{\bf{O}}}
\def\sky{{\theta}}

\def\s{{\bf{s}}}

\def\Ih{{\hat{\I}}}

\def\F{{\bf{F}}}

\def\m{{\bf{m}}}

\def\D{{\bf{D}^{-1}}}
\def\G{{\bf{G}}}
\def\Ni{{\N}^{-1}}
\def\L{{\bf{L}}}
\def\H{{\bf{H}}}

\title[Software Holography for Interferometry]{Software Holography: Interferometric Data Analysis for the Challenges of Next Generation Observatories}
\author[M. F. Morales and M. Matejek]
  {Miguel F. Morales,$^{1,2}$ and Michael Matejek$^{2}$\\
  $^1$University of Washington, Seattle\\
  $^{2}$MIT Kavli Institute}
\date{Released 2008 Xxxxx XX}

\pagerange{\pageref{firstpage}--\pageref{lastpage}} \pubyear{2008}

\begin{document}

\label{firstpage}

\maketitle

\begin{abstract}
Next generation radio observatories such as the MWA, LWA, LOFAR, CARMA and SKA provide a number of challenges for interferometric data analysis. These challenges include heterogeneous arrays, direction-dependent instrumental gain, and refractive and scintillating atmospheric conditions. From the analysis perspective, this means that calibration solutions can not be described using a single complex gain per antenna. In this paper we use the optimal map-making formalism developed for CMB analyses to extend traditional interferometric radio analysis techniques---removing the assumption of a single complex gain per antenna and allowing more complete descriptions of the instrumental and atmospheric conditions.  Due to the similarity with holographic mapping of radio antenna surfaces, we call this extended analysis approach software holography. The resulting analysis algorithms are computationally efficient, unbiased, and optimally sensitive. We show how software holography can be used to solve some of the challenges of next generation observations, and how more familiar analysis techniques can be derived as limiting cases.
\end{abstract}

\section{Introduction}

Motivated by the requirements of next generation radio observatories, we examine an alternative approach for formulating optimal radio analyses. The CMB optimal map-making (OMM) formalism provides an elegant way to translate from a mathematical description of the measurement to provably optimal analyses \citep{Tegmark:1997p2009,Tegmark:1997p2012}, and underlies most current CMB observations. In this paper we use this optimal map-making formalism to describe the interferometric data analysis problem, and show how many of the issues faced by next generation arrays can be naturally incorporated into this framework.

After briefly introducing optimal map making and how it can be used with interferometric data in \S\ref{OMMsec}, we use this framework to expand the mathematical descriptions to include direction-dependent antenna response (\S\ref{InsCalSec}), heterogeneous arrays (\S\ref{heteroArrays}), widefield refractive atmospheric distortions (\S\ref{regime3}), and scintillating distortions (\S\ref{regime4}). We then conclude in \S\ref{Conclusion} with a discussion of widefield effects and comparing this analysis approach with faceting and other commonly used techniques for analyzing interferometric data with direction-dependent calibration and distortion.


In a recent paper \cite{Bhatnagar:2008p2333} detail a new analysis for data with direction-dependent antenna gains which is functionally identical to the analysis we develop in \S \ref{heteroArrays}. While these papers were developed independently, we believe they should be considered as a complimentary pair---Bhatnagar et al. demonstrate increased fidelity in the context of traditional radio astronomy software, while we provide a theoretical foundation for the software holography technique and extend it to a number of other problems facing next generation interferometric arrays.  Specifically, in this paper we will:
\begin{itemize}
  \item Place the work of  \cite{Bhatnagar:2008p2333} on a firm theoretical foundation.
  \item Extend the ideas of software holography to refractive and scintillating atmospheric distortions.
  \item Provide a first step towards using CMB deconvolution techniques with interferometric data, enabling high-precision statistical measurements such as 21 cm Epoch of Reionization power spectrum measurements.
\end{itemize}

\section{Optimal Map Making}
\label{OMMsec}

In this section we briefly introduce the optimal map-making formalism (OMM) that underpins most CMB data analyses \citep{Tegmark:1997p2009,Tegmark:1997p2012}, then as an example show how this description can be used to describe the traditional algorithms used in radio astronomy analysis software such as AIPS, MIRIAD, and CASA.  [Throughout we use linear algebra notation as it allows the expressions to be more compact and expressive. Table \ref{rosetta} and Appendix \ref{transApp} allow the equations in this paper to be converted to integral equivalents, and Appendix \ref{LAintro} provides a brief refresher on linear algebra notation and concepts.]

There are two key steps in the OMM method:  a mathematical description of the measurement, and the optimal reconstruction based on this measurement description. In general, one describes the observation in the following form:
\beq
\label{EqMinBase}
\d = \M \x + \n
\eeq
where $\d$ is a vector of the measurements and $\x$ is the true values one is measuring. $\M$ is then a matrix operator that describes the measurement process, including all instrumental, atmospheric, and data handling effects, and $\n$ is detector noise with covariance matrix $\N \equiv \n\n^{T}$. 

If we assume the measurement can be expressed as linear equation of the form in Eq. \ref{EqMinBase} and the noise is Gaussian and uncorrelated with the signal (both are true for radio astronomy), it can then be proved that the minimally biased estimator for $\x$ is given by
\beq
\hx = (\M^{T} \N^{-1} \M)^{-1} \M^{T} \N^{-1} \d
\label{optEstEq}
\eeq
\citep{Tegmark:1997p2009}. Equation \ref{optEstEq} can be viewed as consisting of two separate parts
\beq
\label{optEstStepseq}
\hx = \underbrace{\left[ \M^{T}\N^{-1}\M\right]^{-1}}_{2} \underbrace{\M^{T}\N^{-1}\d}_{1}. 
\eeq
In the first step the measurements are weighted by their signal-to-noise (including covariant noise) and translated by the conjugate transpose of the measurement description $\M$ back into the coordinates of the input sky. Effectively this forms a `dirty map' at the end of step one. The second part of Equation \ref{optEstStepseq} then represents a deconvolution step. While the OMM method implies a particular style of deconvolution that has been very successful for CMB analysis, one could also use CLEAN, MEM or other non-linear deconvolution algorithms.

For our case, we are only concerned with the first part of Equation \ref{optEstStepseq}, and can thus rewrite the relationship as
\beq
\hx = \D\ \M^{T} \N^{-1} \d,
\label{optEstEqD}
\eeq
where $\D$ represents the deconvolution algorithm of the reader's choice. What is powerful about the OMM framework is that the dirty map formed by the first part of Equation \ref{optEstStepseq} can be proved to be unbiased, lossless, and efficient \citep{Tegmark:1997p2009}. In particular the lossless nature guarantees that all of the information present in the individual measurements $\d$ is retained in the `dirty map' formed at the end of part 1.\footnote{Lossless here means that all of the sky information that was in the visibilities is preserved in the intermediate map. This \emph{does not} mean one can obtain the true sky, as the measurement process itself removes a lot of information (incomplete $u,v$ coverage, etc.), only that the information content of the measurements has been preserved in the analysis. } Often there is significant data compression in forming this lossless intermediate representation. For example, in CMB satellites the huge number of time-ordered-data values are reduced to an unprocessed temperature map, and for VLA interferometry the raw visibilities are reduced to a $u,v$ grid or a `dirty map.' As long as our description of the measurement ($\M$) is accurate none of the information in the raw measurements has been lost in this step.

\subsection{Standard Interferometric Techniques}

It is instructive at this point to write down the standard interferometric analysis of AIPS, MIRIAD, and CASA using the optimal map-making formalism. It is traditional to assume a small field-of-view when deriving the interferometric analysis equations (e.g. \citet{SynthImagIIch1} equation 1--8), and we make the same assumption here as it simplifies the notation in the following sections and helps focus the reader on the unique characteristics of the software holography approach. However, we realize this is a bit of a strawman comparison as there are more advanced techniques in general usage. In \S\ref{Conclusion} we will return to show how widefield effects can be included in all of the developments presented here and discuss how this work is related to the more modern approaches of \citet{Cornwell:1992p2458}, \citet{Sault:1996p2081} and \citet{Bhatnagar:2008p2333}.

We start with a description of the measurement, for a standard mid-frequency observation with an array like the VLA:
\beq
\m(\v) = \G(\v,\v)\Sv(\v,\u)\F(\u,\sky)I(\sky) + \n(\v).
\label{VLAesqObs}
\eeq
(Please see Appendix \ref{transApp} for how to translate this into integral notation.) In words the measurement equation takes the true sky $I(\sky)$ as a vector, Fourier transforms this to form the true $u,v$ distribution (represented by the two-dimensional vector $\u$, see Table \ref{rosetta}), samples the true $u,v$ distribution with the baseline distribution of the observation $\Sv$ (a set of $\delta$-functions at each baseline location) to form the visibilities, multiplies by the complex gain $\G$ appropriate for each visibility (can include both instrumental and atmospheric/ionospheric effects), and adds the per visibility thermal noise $\n$ (usually assumed to be independent for each visibility, but formally cross talk and co-variance can be included). 

Using this description of the measurement, we can directly write down the optimal analysis as
\beq	
\Ih(\sky) = \D\ \F^{T}(\sky,\u)\Sv^{T}(\u,\v) \left[ \G^{T}(\v,\v)\Ni \right] \m(\v).
\label{StandRec}
\eeq
The analysis is essentially weighting by the noise, applying the steps describing the measurement in reverse order (and conjugate transposed), and deconvolving. Again describing the process:  we start with a vector of measured visibilities $\m$, weight them by the inverse noise co-variance matrix (high noise channels receive less weight), multiply by the transpose of the gain $\G^{T}$ to correct the phase, then grid the visibilities to the $u,v$ plane with $S^{T}$ and Fourier transform to form a dirty map of the sky.\footnote{Note that the operator argument $(\u,\v)$ refers to mapping from visibilities ($\v$) to the $u,v$ plane ($\u$). Please see Table \ref{rosetta} for details.}$^{\rm ,}$\footnote{In some radio software implementations the conjugate reciprocal of the gain is used ($1/g_v^*$ or $1/\G^{T}$) as opposed to the gain conjugate ($g_v^*$ or $\G^T$) as depicted in Equation \ref{StandRec}. This difference is usually unimportant in a non-linear CLEAN like algorithm (other than the units of the intermediate map), but the version in Equation \ref{StandRec} maximizes the signal-to-noise. }

Equation \ref{StandRec} is the traditional analysis as implemented in several current interferometric software packages. Often the terms in square brackets are combined to form a single $T_{\rm sys}$ weighting and calibration step using the results of self-cal; sometimes the Fourier transform is incorporated into the deconvolution step for computational reasons; and if a Fast Fourier Transform is used anti-aliasing filters must be added. However, the fundamental algorithm is the same. This is reassuring as it has long been known that this algorithm is optimal \emph{if the description of the measurement in Equation \ref{VLAesqObs} holds}.

The problem encountered by next generation arrays is that Equation \ref{VLAesqObs} does not accurately describe their measurement---there are a number of assumptions about the measurement embedded in this measurement description. Describing the gain as a per-visibility complex number $\G(\v,\v)$ assumes that the gain and phase are uniform across the antenna field of view. Several next generation instruments have instrumental gain and phase which varies as a function of direction within the field of view, fundamentally breaking this assumption. Similarly atmospheric distortions with length scales smaller than the field of view cannot be expressed as a single per-baseline complex number. In addition, the flat sky assumption is incorporated in the $\sky$ coordinate system, hampering the analysis of widefield observations. 

The remainder of this paper largely consists of rewriting Equation \ref{VLAesqObs} to accurately describe the measurements proposed with next generation arrays, and using the optimal map-making formalism to determine the appropriate analysis methods. Throughout this paper we assume the antenna calibration has been determined separately. In current software the overall antenna and atmospheric delay is usually measured via self-cal as part of the deconvolution process, but polarimetric calibration is usually determined through separate parallactic or holographic observations. Either approach to determining the calibration can be used in this formalism.

\section{Position Dependent Calibration \& Heterogeneous Arrays}
\label{InsCalSec}

This section concentrates on instrumental calibration when the gain and phase vary as a function of direction within the field of view. This case is commonly encountered in widefield imaging, such as low frequency observations with the VLA and all upcoming low frequency arrays such as the MWA, LWA, and LOFAR. The additional complication of atmospheric calibration will be delayed until Section \ref{IonoSec}.

We will first assume that the gain pattern is identical for all antennas. The standard measurement description in Equation \ref{VLAesqObs} can be modified to form
\beq
\label{MBeamImageEq}
 \m(\v) = \Sv(\v,\u)\F(\u,\sky)\B(\sky,\sky)I(\sky) + \n(\v).
\eeq
We have replaced the per baseline gain $\G(\v,\v)$ with a direction dependent complex power pattern $\B(\sky,\sky)$. The beam pattern attenuates the signal seen by the interferometer, but the remainder of the measurement is unaffected. 

Again following the OMM framework our analysis method should be
 \beq
 \label{RBeamImageEq}
 \Ih(\sky) =\D\ \B^{T}(\sky,\sky)\F^{T}(\sky,\u) \Sv^{T}(\u,\v)\N^{-1}\m(\v).
 \eeq
This is largely identical to Equation \ref{StandRec}, except that we have multiplied by the beam transpose $\B^{T}$. In forming this transpose, we take the complex conjugate of the gain towards each sky pixel $\sky$ and reorder the entries. This means that a pixel gain of $\frac{1}{2}e^{+i\phi}$ becomes $\frac{1}{2}e^{-i\phi}$:  the phase is corrected but the gain amplitude is applied a second time. The amplitude of the `dirty map' formed just before the deconvolution is attenuated by the beam shape squared. The signal is attenuated once in the measurement description ($\B$ in Eq. \ref{MBeamImageEq}), and again by the analysis procedure ($\B^{T}$ in Eq. \ref{RBeamImageEq}). 
 
While puzzling at first, this beam squared weighting is necessary to form an optimal map. The highest signal-to-noise is achieved if signals are variance weighted, and the measured signal-to-noise is given by the beam pattern. The NVSS team uses beam-squared weighting to add overlapping maps for exactly this reason \citep{Condon:1998p2334}, and it is gratifying to see this result fall out of this derivation. Of course the deconvolution algorithm must understand the weighting of the dirty map to appropriately reconstruct a final image.

\subsection{Heterogeneous Arrays}
\label{heteroArrays}

The development above assumed that the directional response of all the antennas were identical. For many upcoming observations the antennas are not identical, either due to antenna-to-antenna variation such as the MWA, or due to a mix of different antenna types as in CARMA. In this section we will remove the assumption of identical antenna responses, and follow a slightly more detailed derivation to illustrate use of the OMM method. 

The easiest way to extend Equation \ref{MBeamImageEq} to heterogeneous arrays is to subscript the beam pattern, giving each baseline a unique power pattern
\beq
\label{MBeamImage2Eq}
 \m(\v) = \Sv(\v,\u_{b})\F(\u_{b},\sky_{b})\B_{b}(\sky_{b},\sky)I(\sky) + \n(\v).
\eeq
This equation creates a different observed sky for each antenna pair $\sky_{b}$, and the remainder of the measurement remains the same. Unfortunately this is a computationally expensive description of the measurement as each of the $b$ separate observed skies require a Fourier transform and sampling, and it leads to a computationally expensive analysis algorithm:
 \beq
 \label{RBeamImage2Eq}
 \Ih(\sky) =\D\ \B^{T}(\sky,\sky_{b})\F^{T}(\sky_{b},\u_{b}) \Sv^{T}(\u_{b},\v)\N^{-1}\m(\v).
 \eeq
Here each baseline is gridded (single $u,v$ $\delta$-function) and Fourier transformed to produce a single fringe, which is then attenuated by the power pattern appropriate for that baseline before being added to a common dirty map. Conceptually, this is correct. The fringe should only be added to portions of the image seen by that antenna pair, and because $\B$ is complex the location of fringe peaks can shift from one portion of the image to another in response to direction dependent phase response.

Fortunately, there is a more efficient way to perform the same analysis. Returning to the measurement description in Equation \ref{MBeamImage2Eq}, we can recast the problem
\beqa
\m(\v) & = &\Sv(\v,\u_{b})\F(\u_{b},\sky_{b})\B_{b}(\sky_{b},\sky)I(\sky) + \n(\v), \nonumber \\
\m(\v) & = &\Sv(\v,\u_{b})\B_{b}(\u_{b},\u)\F(\u,\sky)I(\sky) + \n(\v), \nonumber \\
\m(\v) & = &\B(\v,\u)\F(\u,\sky)I(\sky) + \n(\v).
\label{MBeamEq}
\eeqa
In going from line 1 to 2, we have simply pulled the power pattern $\B$ to the other side of Fourier transform and expressed the power response in $u,v$ coordinates (the operators commute because the Fourier transform is unitary, and $\B_{b}$ becomes a convolution). The beam pattern is still baseline dependent in line 2, but the sampling function $\Sv$ is already selecting out individual baselines to create visibility measurements, so these operators can be combined in the last line.

In words, the input sky is transformed to $u,v$ coordinates, then the appropriate region of the $u,v$ plane is integrated to form the visibility measured by that pair of antennas using the unique power response of that antenna pair. It is interesting to note that Equation \ref{MBeamEq} is the algorithm used for simulating the response of interferometric observations by the MIT Array Performance Simulator (MAPS).

Before moving to the analysis, a natural question is the size of the power pattern $\B$ in $u,v$ coordinates---if it covers a significant portion of the $u,v$ plane it would remain computationally expensive. Each antenna has a response to the incident electric field $\W_{a}(\sky,\sky)$, where the response is complex to capture both the electric field gain and phase delay. We can transform the antenna response to $u,v$ coordinates $\W_{a}(\u,\u)$. This transformed antenna response is exactly what is obtained during a holographic measurement of the antenna gain \citep{Scott:1977p2425}. Because the electric field outside the antenna physically cannot be added into the received signal, $\W_{a}(\u,\u)$ has the same size as the antenna.\footnote{Reflections from outside the antenna and widefield w-projection effects can make the response slightly larger, but it remains very compact in the $u,v$ plane.} The power pattern observed by a baseline is given by the multiplication of the constituent antenna sky responses, or the convolution of the $u,v$ plane responses of antennas $i$ and $j$
\beq
\B_{ij}(\u,\u) = \W_{i}^{T}(\u,\u)\ast\W_{j}(\u,\u).
\label{Beamuv}
\eeq
Thus the power pattern $\B(u,u)$ is very compact in the $u,v$ plane, approximately twice the physical width of an antenna. This compact feature is why it forms the basis of array simulators.

Moving to the analysis, we can again use OMM to form an optimal analysis approach:
 \beq
 \label{RBeamEq}
 \Ih(\sky) =\D\ \F^{T}(\sky,\u) \B^{T}(\u,\v) \N^{-1}\m(\v).
 \eeq
 In this analysis we have replaced the simple $\delta$-function gridding of $\Sv^{T}(\u,\v)$ with a gridding function $\B^{T}(\u,\v)$ that spreads the visibility out on the $u,v$ plane using the power pattern response of that particular antenna pair. Effectively the direction and baseline dependent instrumental calibration has become part of the gridding kernel.\footnote{When using an FFT, an anti-aliasing filter must be added to the power response kernel using an additional convolution. This has no impact on the final precision if the effects of the anti-aliasing filter are included in the deconvolution.} Because this uses the holographic antenna response to perform the calibration, we call this analysis technique software holography.

\begin{figure*}
\begin{center}
\includegraphics[width = 12 cm]{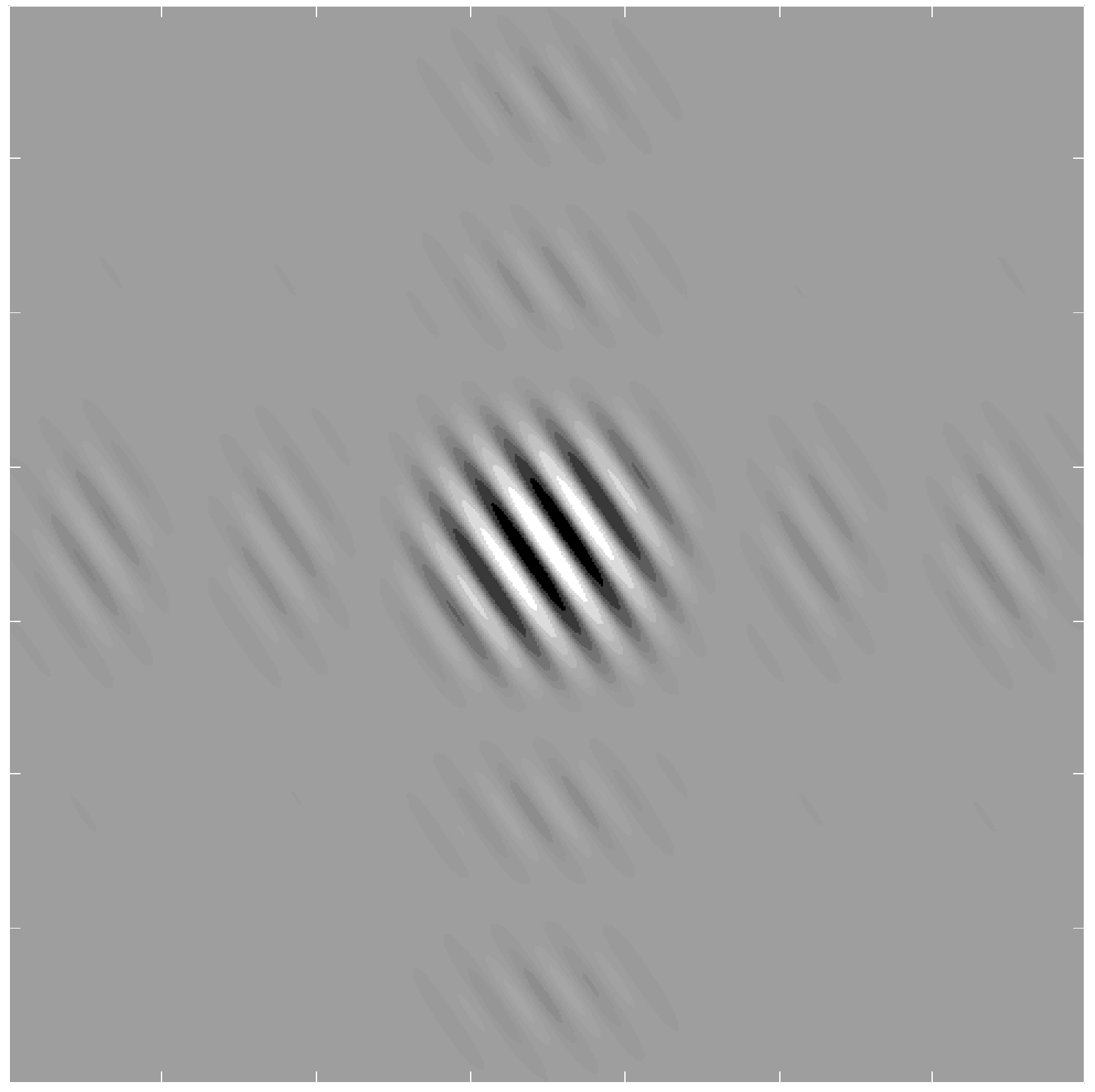}
\caption{This figure shows the contours from a dirty map for a single baseline using software holography. For this example we have used an antenna consisting of a 4x4 array of dipoles with a spacing of nearly one wavelength so there are strong grating lobes, and a very widefield image to show both the primary beam (center) and grating sidelobes (surrounding). This is similar to the beam patterns seen by LOFAR and MWA antennas near the top of their frequency bands. In the image the fringe from the single visibility is clearly seen as the diagonal corrugations, but its amplitude has been enveloped by the known antenna pattern and the sidelobes are clearly evident. While the beam pattern covers the sky, the corresponding convolution in the $u,v$ plane is very compact. In traditional interferometric analysis, the corrugations would have the same amplitude across the image, as the \emph{prior} of the antenna pattern is not used. In software holography the enveloping power pattern can vary from baseline-to-baseline to accurately represent the directional sensitivity of individual antenna pairs. Not shown here is the direction-dependent shifting of the fringe peaks which can be produced by directional differences in the phase delays of the antennas.}
\label{MBEmap}
\end{center}
\end{figure*}

From a statistical viewpoint software holography can be understood as adding a Bayesian prior to the analysis. In the standard analysis (Equation \ref{StandRec}) it is assumed that the antennas have uniform power responses across the image, and the calibration simply affects the amplitude and phase of the fringe measured by a single baseline. In the software holography analysis, we are adding the prior that we know the direction-dependent gain of each antenna. With this prior, only the portions of the sky the given antenna pair were sensitive to should be reconstructed with a fringe---if the antennas could not respond to radiation from a portion of the sky, the measured visibility should not be interpreted as coming from that direction. This enveloping of the fringe by the baseline power response is shown graphically in Figure \ref{MBEmap}.

Not including the antenna-dependent holographic measurements (when known) is a form of bias. Incorporating these effects using software holography improves the precision of high-dynamic range measurements for current observations with the VLA and similar interferometers \citep{Bhatnagar:2008p2333}, in addition to observations with next generation instruments.


\section{Atmospheric \& Ionospheric Distortions}
\label{IonoSec}

\renewcommand{\labelenumi}{\arabic{enumi})}
\cite{LonsdaleIonosphere} provides a review of atmospheric and ionospheric distortions. In that work the effects of atmospheric disturbance are separated into four regimes, depending on the characteristics of the interferometer and the length scales of atmospheric disturbances. Briefly these regimes are:

\begin{enumerate}
  \item A narrow field of view and short baselines. Distortion appears as a translation of entire field, correctable with a tip-tilt compensation.
  \item A narrow field of view and long baselines. Independent phase delay for each antenna, but the delay applies to entire antenna field of view. Appears as scintillation, but same scintillation pattern for all sources in the field. Correctable with single calibrator adaptive optics and self-cal, and is typical for VLA observations.
  \item Short baselines but a wide field of view. Distortion is purely refractive, but varies across the field of view. Sometimes described as a rubber-sheet distortion and is typical of MWA observations
  \item Wide field of view and long baselines. The worst case, as sources scintillate across the field of view with a position dependent scintillation screen. This is the challenge faced by the LWA and LOFAR at the longest baselines.
\end{enumerate}

Regimes 1 \& 2 are described by the standard measurement description (Equation \ref{VLAesqObs}), and are well handled with current analysis algorithms. Regimes 3 \& 4 cannot be described as a single phase delay per antenna, because the delay varies across the field of view. In the following sections we will explore how to use OMM to analyze observations in the challenging atmospheric conditions or regimes 3 \& 4.

\subsection{Widefield Refractive Distortions}
\label{regime3}

In regime 3, atmospheric and ionospheric distortions appear as a rubber-sheet distortion:  the apparent positions of sources are shifted but they do not appear to scintillate. Mathematically this can be described by:
\beq
\label{Regime3}
 \m(\v) = \B(\v,\u_{t})\F(\u_{t},\sky'_{t})\A(\sky'_{t},\sky; t)I(\sky) + \n(\v').
\eeq
Here we have added a time-dependent atmospheric distortion $\A(\sky'_{t},\sky; t)$ that moves the apparent locations of the sources seen by the array from $\sky$ to $\sky'$. The atmospheric distortion presented here can also include position dependent absorption and Faraday rotation, as long as the distortion is the same for all antennas in the array. Because the atmospheric distortion is time-dependent, the coordinate mapping from the real sky to the apparent sky changes. This means that the associated $u,v$ coordinates and visibilities are also time dependent as indicated. We have also included baseline dependent antenna calibration ($\B$) in this example to illustrate how effects can be stacked. 

The corresponding analysis is
 \beq
 \label{Regime3anal}
 \Ih(\sky) =\D\ \A^{T}(\sky,\sky'_{t}; t)\F^{T}(\sky'_{t},\u_{t}) \B^{T}(\u_{t},\v) \N^{-1}\m(\v).
 \eeq
Through the Fourier transform, this is identical to the analysis in the previous section. However, because of the time-dependent nature of the atmospheric distortion we can only grid and Fourier transform visibilities from one atmospheric realization. Effectively we are creating instrumentally calibrated snapshot images of the apparent sky, which we then correct with a rubber-sheet correction so sources appear in their true locations ($\sky$), and then stack the snapshot images. The time scale of the snapshot images is set by the atmospheric distortions. This analysis is effectively the approach taken by the MWA \citep{Mitchell:2007p144}, and the ionospheric distortion timescale sets the 8 second snapshot cadence of the instrument.

It is possible to Fourier transform the atmospheric operator and describe it in the $u,v$ plane, in analogy to what we did with the antenna power response in Section \ref{heteroArrays}. We will look at this approach in the context of the next section, but the snapshot imaging appears to be a better solution for most interferometers operating in the rubber-sheet conditions considered here.

\subsection{Widefield Scintillating Distortions}
\label{regime4}

Scintillating widefield distortions of regime 4 are the most difficult, because each antenna sees a different atmospheric phase screen. This is the analysis challenge faced by LOFAR and the LWA.

To describe the widefield scintillation seen in regime 4 we need to make our atmosphere model more general. The wavefront observed by an individual antenna can be described as a direction-dependent phase delay $\L_{a}(\sky,\sky; t)$ (labeled $\L$ in sympathy with the challenges faced by LOFAR and the LWA). The atmospheric distortion seen by a baseline is then given by 
\beqa
\A_{ij}(\sky,\sky) &=& \L_{i}^{T}(\sky,\sky)\L_{j}(\sky,\sky), \label{Asky}\\
\A_{ij}(\u,\u) &=& \L_{i}^{T}(\u,\u)\ast\L_{j}(\u,\u). \label{Auv}
\eeqa
Effectively $\A_{b}$ is describing the distortion to the fringe pattern for that baseline. Under the regime 4 atmospheric conditions the fringe pattern for a single visibility will not appear as a simple grating across the field, but more like the lines on a topographic map as the position dependent ionospheric delay shifts and distorts the locations of the fringe peaks.

We can use Equations \ref{Asky} and \ref{Auv} to describe the measurement as
\beq
\label{Regime4}
 \m(\v) = \B(\v,\u_{t,b})\F(\u_{t,b},\sky'_{t,b})\A_{b}(\sky'_{t,b},\sky; t)I(\sky) + \n(\v').
\eeq
Not only is the distortion time variable, it is different for every baseline. This leads to an analysis of the form
 \beq
 \label{Regime4anal}
 \Ih(\sky) =\D\ \A^{T}(\sky,\sky_{t,b}; t)\F^{T}(\sky_{t,b},\u_{t,b}) \B^{T}(\u_{t,b},\v) \N^{-1}\m(\v).
 \eeq
 This analysis is very computationally expensive. In words, we must Fourier transform each calibrated visibility to create a snapshot image of \emph{that} fringe, which is then corrected by the baseline dependent atmospheric operator $\A^{T}$. The individual fringes are then co-added to form a snapshot image, and successive images are stacked to form an integrated dirty map.

As an alternative, we could Fourier transform the atmospheric distortion and pull it to the left of the Fourier transform in analogy to Equation \ref{MBeamEq} to obtain
\beq
\label{Regime4uv}
 \m(\v) = \big[\B(\v,\u_{t,b})\A(\u_{t,b},\u; t)\big]\F(\u,\sky)I(\sky) + \n(\v').
\eeq
As implied by the square brackets, we can now envision combining the position dependent atmospheric and instrumental distortions into a single $u,v$ plane operation. This would give us the the analysis procedure
\beq
\label{Regime4uvanal}
 \Ih(\sky) =\D\ \F^{T}(\sky,\u) \left[ \A^{T}(\u,\u_{b})\B^{T}(\u_{b},\v)\right] \N^{-1}\m(\v).
\eeq
At first glance this appears to be the obvious solution:  correct both the scintillating ionosphere and instrumental gain in one baseline dependent gridding step. The difficulty is that unlike the beam response, the atmospheric distortion is not well localized in the $u,v$ plane \citep{MatejeckIonoMemo}.

The difference in the locality of the instrumental and atmospheric terms is related to the physics of the two distortions. The electric field response $\W$ of an antenna is fundamentally the sum of the electric field collected at each location on the antenna ($\W = w_{1} + w_{2} + ....$). These terms are complex, and if they are added out of phase they interfere and destroy the response of the antenna. The atmospheric delay $\L$ on the other hand is multiplicative:  if we decompose the atmosphere into many terms each rotates the phase by a certain delay angle $d$. Mathematically this ends up with a form of $e^{2\pi i (d_{1}+d_{2}+... )}$. Due to the multiplicative nature, this is not compact in the $u,v$ plane and strong beating effects between atmospheric modes come into play---the spatial equivalent of intermodulation distortion. We refer the interested reader to \cite{MatejeckIonoMemo}.

In conclusion, if the ionospheric phase screen is just a little more complicated than a single per-antenna delay (regime 2) but can be described with only a couple of large sinusoidal modes for each antenna (barely into regime 4), the $u,v$ plane atmospheric correction described in Equation \ref{Regime4uvanal} may be useful. However, for complex direction and antenna dependent atmospheres the $u,v$ size of the correction becomes enormous. Under these most challenging conditions one would be best served with the conceptually simpler snapshot fringe approach of Equation \ref{Regime4anal}.

\section{Discussion}
\label{Conclusion}

To fully integrate software holography into modern interferometric data analysis there are a few loose ends we should tie up, including projection effects and a comparison with multi-faceting techniques.

So far we have used a flat sky ($\sky$) and simple Fourier relationship to simplify the notation and help focus on the unique characteristics of software holography. In general this is a poor assumption, particularly in the context of widefield atmospheric distortions. Fortunately widefield/w-projection effects can be easily added. 

Returning to basics and following the discussion in lecture 1 of Synthesis Imaging in Radio Astronomy \citep{SynthImagIIch1}, the general spatial correlation relationship can be written as (their Equation 1-5)
\beq
{\bf V}(u,v,w) ={\bf C}(\{u,v,w\},\s) I(\s),
\eeq	
where
\beq
{\bf C}(\{u,v,w\},\s) = \int e^{-2\pi i \nu \s \cdot \w/c} d^{2}s
\eeq
and $\w = \{u,v,w\} = {\bf r_{1}} - {\bf r_{2}}$.
Following \citet{Cornwell:2003p2045,Cornwell:2008p3151}, in  $\{u,v,w\}$ coordinates the correlation relation ${\bf C}$ can be decomposed into a Fourier transform, a coordinate conversion $\T$ from $\s \rightarrow \{l,m\}$, and an additional term $\H$ (their Equation 10)
\beq
{\bf C} = \F\big(\{u,v\},\{l,m\}\big)\H\big(\{l,m,w\},\{l,m\}\big)\T\big(\{l,m\},\s\big),
\eeq
where $\H = e^{-2\pi i [w(\sqrt{1 - l^{2}-m^{2}} -1)]}.$
This gives us the three standard limiting cases:
\begin{itemize}
  \item If the field of view is small, $\H$ is negligible and can be ignored.
  \item If the array is coplanar, $\H$ can be kept negligible at the cost of a time-dependent coordinate conversion $\T(\{l',m'\},\s)$. For observatories that contend with a direction-dependent atmospheric refraction (\S \ref{regime3}), this coordinate transformation can be combined with the time-dependent atmospheric distortion $\A(\sky',\sky; t)$ and corrected at no additional cost.
  \item In the most general case, we can follow the w-projection technique developed by \citet{Cornwell:2003p2045,Cornwell:2008p3151}. This is equivalent to pulling $\H$ through the Fourier transform to create a widefield $u,v$ correction $\H(\u,\u;w,t)$. 
\end{itemize}
Any of these limits can be added to the analyses developed in this paper by inserting the appropriate operators. 

As an aside, Cornwell et al.\ worked around the pole in $\H(\u,\u)$ by using the anti-aliasing filter in the gridding step, even though this filter is only needed for Fast Fourier Transforms. In the software-holography context this can be more physically interpreted as a convolution with the holographic beam pattern $\B(\u,\u)$. Since any real antenna has a non-zero size, the pole in $\H$ naturally goes away for any physical system.

All of the analysis issues approached in this paper have been successfully tackled in the past with mosaic imaging and multi-faceting deconvolution techniques of \citet{Cornwell:1992p2458} and \citet{Sault:1996p2081}. These techniques subdivide the field-of-view and apply a separate complex gain calibration to each facet, allowing the atmospheric and instrumental calibration to change across the field in a stepwise fashion. Faceting has the definite advantage of being a proven technique behind many astronomy results, but for next generation arrays its requirements on data storage and computational efficiency may make software holography an attractive alternative.

The computational requirements of faceting are driven by how faceting is integrated into the deconvolution process. When a faceted dirty image is produced, the facet edges distort the apparent array beam of sources in neighboring facets. This is typically dealt with by using the dirty map only as an intermediate step in the deconvolution process---visibilities are used to make a faceted map, sources are identified and subtracted from the raw visibility data, and a new faceted dirty map is created for the subsequent iteration of the deconvolution algorithm. The deconvolution algorithm must always work on the full visibility dataset, as the dirty image contains artifacts which cannot be easily removed.

\citet{Bhatnagar:2008p2333} have recently demonstrated the use of software holography within a traditional non-linear deconvolution algorithm. Their algorithm is significantly faster than faceting and does not suffer the discontinuities and artifacts in the intermediate image. However, their algorithm still subtracts the sky model from the raw visibilities, using the dirty map only as an intermediate step in the deconvolution process.

For next generation radio arrays with hundreds of elements and wide fields of view the raw visibility data can be very large: for example the correlated data rate for the MWA is $\sim$19 GBytes/s over just 31 MHz of bandwidth. The optimal map making technique was developed in response to these same computational problems as faced by the CMB community \citep{Tegmark:1997p2012}. The time series data from a satellite such as WMAP and Plank is analogous to interferometric visibilities and similarly voluminous, and deconvolution of the time series data quickly becomes computationally impractical. OMM allows all of the information in the raw data measurements to be preserved in the intermediate map, reducing both storage needs and the computational requirements of deconvolution. The precision of CMB measurements is a testament to the power of the optimal map making formalism.

In interferometric software holography, forming the intermediate map is computationally very efficient. The direction-dependent gain of each antenna can be corrected by gridding with the baseline dependent $u,v$ power pattern. This gridding kernel is very compact, leading to a efficient imaging algorithm. The atmospheric corrections are less efficient than the instrumental calibration, but the final map has no significant artifacts. Thus the deconvolution algorithms can work directly on the lossless `dirty map' formed by software holography, without referring to the raw visibilities. Deconvolving intermediate maps should be no slower than deconvolving the raw visibilities (while requiring much less storage space), and potentially could be much faster if techniques from the CMB community can be effectively used.

It is hoped the software holography techniques presented in this paper will assist the development of analysis systems for next generation instrumentation, and enable precision interferometric measurements such as power spectrum detection of 21 cm emission from the Epoch of Reionization.

\section*{Acknowledgements}
There are a number of people which have had significant impacts on this paper, either through in-depth discussions or comments on early drafts. In particular we would like to thank Matias Zaldarriaga, Max Tegmark, Colin Lonsdale, Roger Cappallo, Jacqueline Hewitt, David Kaplan, Randall Wayth, Daniel Mitchell, Bob Sault, and Judd Bowman. This work has been supported by NSF grant \#AST-0457585 and the MIT School of Science.

\bibliographystyle{mn2e}
\bibliography{holographybib,misc}

\appendix

\section{Translating to integral notation}
\label{transApp}

Linear algebra notation is very efficient for the describing measurements and the related optimal analysis procedures. However, much of the interferometry literature is written in integral notation. To aid translation, the key operators in this paper are listed in Table \ref{rosetta} with their integral equivalents and a brief description of each mathematical operation.

As a simple example of how to convert a linear algebra equation into its integral equivalent, the basic measurement description in Equation \ref{VLAesqObs}
\beq
\m(\v) = \G(\v,\v)\Sv(\v,\u)\F(\u,\sky)I(\sky) + \n(\v)
\label{VLAesqObsApp}
\eeq
can be rewritten using Table \ref{rosetta} to create
\beq
m_v = g_v\int \delta(\u-\u_b) \left[ \int e^{-2\pi i \u \cdot \sky}I(\sky)\  d^2 \sky\right]   d^2\u + n_v.
\label{intEx}
\eeq
In creating Equation \ref{intEx} each integral from Table \ref{rosetta} encloses all of the expressions which appear to the right in the linear algebra version. Using this procedure all of the equations in this paper can be translated into integral equivalents.

\begin{table*}
\begin{minipage}{7in}
\caption{This table lists the operators and vectors used in this paper, along with the integral formulation and comments on the operation.}
\label{rosetta}
\begin{tabular}{@{}ll p{4 in}}
Linear Algebra & Integral Notation & Comments\\
\hline
$\u$ & $\u$ or $\{u,v\}$ & $u,v$ coordinates. In this paper we condense this to a single two dimensional vector $\u$ to make the notation more compact and avoid confusion with visibilities.\\
$\v$ & $_v$ & `visibility' coordinates, or a vector listing the visibilities. In integral notation usually indicated as a subscript.\\
\hline
$\n(\v)$ & $n_v$ & Thermal noise per visibility. The matrix $\N$ is formed by the outer product of two vectors of the thermal noise, and allows correlated noise to be included in the linear algebra notation (e.g.\ cable cross-talk).\\
$\m(\v)$ & $m_v$ or $v_i$ & A vector of measured visibilities. Usually expressed with subscripts in integral notation.\\
$I(\sky)$ & $I(\sky)$ & The true sky brightness distribution. Note that in the linear algebra notation this is a vector of sky locations---thinking of this as a two dimensional `matrix' of values breaks the linear algebra notation (requires all operators to be diagonal).\\
\hline
$\F(\u,\sky)$ & $\int e^{-2\pi i \u \cdot \sky} d^2 \sky$ & Fourier transform. May be replaced with an FFT with the addition of an anti-aliasing filter.\\
$\Sv(\v,\u)$ & $\int \delta(\u-\u_b)d^2\u$ & Sampling function which selects the locations in the $u,v$-plane which are measured by an interferometric baseline $b$ to create a visibility. In the linear algebra notation the result is a vector of the visibilities ($\v$).\\
$\G(\v,\v)$ & $g_v$ & A single complex gain per visibility. The matrix version has entries only along the diagonal.\\
$\B(\sky,\sky)$ & $B(\sky)$  & The power response of a pair of antennas. As both the gain and phase may change with direction for each antenna, this is a complex function. The $\B$ operator is diagonal.\\
$\B(\u,\u)$ & $B(\u)*$ or $\int B(\u'-\u)d^2\u$ & Power response of a pair of antennas in $u,v$ ($\u$) coordinates. The Fourier transform of $B(\sky)$ and includes the convolution created by the translating the multiplication in $\sky$ coordinates $\u$. Due to the physics of antennas the $\B$ operator is sparse (of limited extent in $\u$). \\
$\B(\v,\u)$ & $\int\int \delta(\u'-\u_b'')B_b(\u'-\u)d^2\u\ d^2\u'$ & Power response of the antennas in a particular baseline. Effectively this is a convolution over the sky in $u,v$ coordinates combined with a delta-function to select the baseline sampled by that antenna pair. The result is a vector of visibilities.\\
$\W_i(\u,\u)$ & $W_i(u)*$ or $\int W_i(\u'-\u) d^2\u$ & The electric field response of an antenna in $u,v$ coordinates. This is the holographic antenna pattern, and is the Fourier transform of the direction-dependent gain $W_i(\sky)$.\\
$\H\big(\{l,m,w\},\{l,m\}\big)$ & $e^{-2\pi i [w(\sqrt{1 - l^{2}-m^{2}} -1)]}$ & $w$-projection. For non-coplanar baselines in the narrow field limit can be interpreted as Fresnel diffraction \citep{Cornwell:2008p3151}\\
$\D$ & --- & Deconvolution. There are many styles of deconvolution, many of which are non-linear (cannot be expressed as an integral or linear algebra equation). Any kind of deconvolution can be used with the results presented in this paper.\\
\hline
\end{tabular}
\end{minipage}
\end{table*}

\section{Brief Introduction to Linear Algebra}
\label{LAintro}
The following is designed as a brief refresher of key linear algebra concepts.

In linear algebra, an operator $\O$ transforms an input vector into a new vector. For many cases, this becomes a generalized change of variables where the result is in a different coordinate set than the input. 
\beq
\d(x) = \O(x,y; z) \v(y)
\eeq
In this notation the operator $\O(x,y;z)$ depends on parameters $z$ and transforms the input vector $\v$ in coordinates $y$ to the result $\d$ in a new set of coordinates $x$. This works perfectly well for continuous coordinates as well as discreet, we just have to imagine breaking the continuous coordinates into very small ``pixels."

For a concrete example, let's imagine calculating the sky temperature $T$ for a set of antennas ${\bf a}$, given a position dependent gain $G(\sky)$ and a distribution of sources $I(\sky)$. In the traditional integral formulation the sky temperature of each antenna is given by
\beq
T_a = \int G_a(\sky) I(\sky) d^2\sky.
\label{Teq}
\eeq
This can be as easily written in operator notation as
\beq
\label{AntTempExLAeq}
{\rm T}({\bf a}) = {\bf G}({\bf a},\sky; p) I(\sky),
\eeq
where $p$ is the parameters the antenna gain depends on. Here the operator ${\bf G}$ is performing the integration and explicitly going from the continuous sky coordinates $\sky$ to the discreet per-antenna coordinates ${\bf a}$. One advantage of the operator notation is adding polarization information and Jones matrices is notationally straightforward, we can just add more pixels to $\sky$, one for each polarization. Note that because the operator ${\bf G}$ includes the integration, ${\bf G}$ and $G$ do not have the same units ($G$ is unitless and ${\bf G}$ has units of str or m$^{-2}$).

There are a couple of common misconceptions which can make reading linear algebra equations more difficult.

The first is the difference between the conjugate transpose and inverse of a matrix operator. The conjugate transpose of an operator simply reverses the rows and columns (and takes conjugate if complex), and serves to reverse the direction of the coordinate transformation. So for our example above we can easily create ${\bf G}^T(\sky, {\bf a};p)$ by reversing the order of the indices and conjugating, and perform the following calculation
\beq
\cI'(\sky) =  {\bf G}^T(\sky, {\bf a};p) {\rm T}({\bf a}),
\eeq
to move from antenna temperatures back to sky coordinates. \emph{However, $\cI'(\sky)$ is not equal to the true sky brightness distribution  $\cI(\sky)$  in Equation \ref{AntTempExLAeq}} . While we are back in the correct coordinates, we have not recreated the distribution of sources in the sky. An antenna temperature does not allow reconstruction of the sky distribution, because of the spatial integral in Equations \ref{Teq} \& \ref{AntTempExLAeq} has erased this information. Additionally, while $\cI'$ is in the same coordinates $\sky$ as the original input vector $\cI$, it is not necessarily in the same \emph{units} (units refers to the value scale at a location whereas coordinates refers to the scale of the map axes).  In our example here the operators ${\bf G}$ and ${\bf G}^T$ both have the same units, not inverse units, so $\cI'(\sky)$ does not have the same units as $I(\sky)$ even though the coordinate scales of the images are the same.

To really go back to the input signal we would need to know the inverse operator ${\bf G}^{-1}(\sky, {\bf a};p)$, and as in most cases the inverse operator does not exist for our example due to the sky integral. We can go back to the original coordinates using the transpose, but we will not obtain the same result we started with  unless the operator is unitary. 

The second common misconception is to think of the sky $I(\sky)$ as a matrix-like quantity. While it is tempting to think of the sky as a two dimensional grid, in linear algebra each of the pixels is a separate entry in a long one-dimensional vector. Thinking of the input as a matrix breaks the linear algebra notation and leads to artificial constraints (effectively requiring the operators to be diagonal). A corollary of this is that the matrix operators are formally very large; if our input vector has $N_{pix}$, the operator is $N_{pix}\times N_{pix}$ in size. But this is a formal definition and does not mean that we have $N_{pix}^{2}$ operations to perform in most cases. The amount of computation comes down to how many of the entries in the operator are populated---the sparseness of the operator. The number of populated entries in the operator rows directly determines the number of computations required, and in most of our examples this number is quite small.

An illustrative example is the refractive ionospheric distortion operator $\A(\sky',\sky)$, which shifts the flux from the true sky location $\sky$ to an apparent location $\sky'$. If there is no scintillation, this is a one-to-one mapping described by a delta-function $\delta(\sky-\sky'-d\sky(\sky,t))$, where $d\sky$ is the refractive shift. The entries in the $\A(\sky',\sky)$ operator consists of ones to map the flux from the input pixel to the output pixel, with a single entry per row. The operator is very sparse and easy to apply computationally, however because it is not diagonal it cannot be formulated if the input sky is thought of as a matrix-like quantity.

 \end{document}